\documentclass[%
 reprint,
 amsmath,amssymb,
 aps,
]{revtex4-2}

\usepackage{graphicx}% Include figure files
\usepackage{color} % for text color
\usepackage[version=4]{mhchem} % for using \ce in bibtex 
\usepackage{dcolumn}% Align table columns on decimal point
\usepackage{bm}% bold math
\usepackage[utf8]{inputenc}
\usepackage{float}
\usepackage[mathlines]{lineno}% Enable numbering of text and display math
\begin{document}

\preprint{APS/123-QED}

\title{Defect Modulated Band Modification in Ion Implanted MgO Crystal: Experimental and Ab Initio Calculations}% Force line breaks with \\

\author{Sourav Bhakta}%
 \affiliation{%
 School of Physical Sciences, National Institute of Science Education and Research Bhubaneswar, An OCC of Homi Bhabha National Institute, Jatni, Odisha-752050, India.
}%
\author{Subhadip Pradhan}%
\author{Ashis K. Nandy}%
\affiliation{%
 School of Physical Sciences, National Institute of Science Education and Research Bhubaneswar, An OCC of Homi Bhabha National Institute, Jatni, Odisha-752050, India.
}%
\author{Pratap K. Sahoo}%
\email{pratap.sahoo@niser.ac.in}%
\affiliation{%
School of Physical Sciences, National Institute of Science Education and Research Bhubaneswar, An OCC of Homi Bhabha National Institute, Jatni, Odisha-752050, India.
}%

\date{\today}% It is always \today, today,

\begin{abstract}
Defects creation and annihilation is a fundamental concept in device fabrication. This report studies the optical bandgap modification in MgO by MeV Ni ion irradiation-induced defect states between valance and conduction band. Ion implantation on MgO single crystal produces substitutional defect states along with F (anionic vacancy center), $F_2$, other oxygen vacancy center and V (cationic vacancy center) centers confirmed from absorption and photoluminescence spectra that can be applied as filament in valance charge memory based resistive random access memories. The variation of optical bandgap with Ni ion fluences is ascertained by modifying the electronic band structure. Density Functional Theory (DFT) calculation assists in understanding the evolution of electronic band structure for vacancies and substitutional defects consisting of MgO structures.

\end{abstract}

\maketitle

%\tableofcontents

\section{\label{sec:level1}Introduction }
Ion implantation is one of the most reliable and novel technique to generate various type of defects in host lattice \cite{1,Markevich2006,3} that can modulate optical \cite{4,5}, magnetic \cite{6}, electronic \cite{7,8} etc. properties. Defect creation in MgO apart from the conventional chemical routes \cite{9,10,11,12} for determining various optical properties applied in different technological applications such as adsorbents, sensors, catalysis, refractory material, paint, fluoride remover, optoelectronics, and luminescence devices \cite{13,14,15,16}. Ionic model of solid and interatomic potentials based modified electron-gas model suggest that each oxygen site incorporated with effective 2- electronic charge compensated by the nearest Mg site associated with effective 2+ electronic charge in MgO\cite{18}. Depending on the impurity ion incorporation in MgO, the vacancies and defects are of different kinds. As the impurity concentration increases, the enthalpy of the system becomes less negative due to the increase of the lattice strain by atomic displacement from its original sites. When MeV Ni ions are implanted in crystalline MgO, a large number of vacancy defects are created by removing O or Mg atom from the host lattice, and these vacancy centers are stable and neutral with respect to charge \cite{19}.\\
The optical properties of MgO depend on the kind of defect formation and their position in lattice. The dramatic change of optical properties can be tuned by the formation of small metallic clusters on the surface of the lattice, which absorbs light in different frequency limits \cite{4}. Ion implantation on MgO surface plays a crucial role in modifying the surface and enhancing electrical and optical properties. The creation of quantum antidots and the modification of the refractive index of MgO by Au ion implantation has already been observed and analyzed \cite{22,23}. The enhancement of photoluminescence property of MgO by  He, Ar, Fe, Cr ion implantation \cite{24,27} and the exhibition of giant magneto-resistance, super-paramagnetism \cite{28,30,31} and ferromagnetic ordering \cite{6,33} by transition metal ion like Fe, Cr, Ni, Co are a good example of such kind of surface modification of MgO. The correlation between defect induced electronic states inside the bandgap and the optical properties of MgO helps to investigate the nature of defect and color centers \cite{34}. The radiation effect via surface modification using 200 keV and 1 MeV Ni ion implantation in MgO single crystal has also been investigated by Mitamura et al. \cite{37}. However, the correlation between the nature of defect centers and the electronic band structure by MeV Ni ion implantation with fluences has not been investigated extensively.       

This report focus on the dependence of defect centers on bandgap variation as a function of Ni ion fluence and origin of light emission with respect to F and V color centers in MgO.  The experimental bandgap and interband transitions due to various defect centers were verified using the electronic structure calculation in the framework of density functional theory by implemented in the Vienna Abinitio Simulation Package. 

%{In this paper, we investigated the induction of different kind of defect states in crystalline MgO by MeV Ni ion implantation to tune the bandgap, and the role of these defect bands are examined theoretically by DFT.}

\section{\label{sec:level2}Experimental Procedure}

Single crystalline MgO (100) substrate of thickness 0.5 mm were brought from the MTI corporation USA. The polished surface were irradiated by 1 MeV Ni$^+$ ions at room temperature at a fluence range from $5\times 10^{14}$ ions/cm$^2$ to $1\times 10^{16}$ ions/cm$^2$ using 3 MV Pelletron Accelerator (NEC, USA) facility at IOP, Bhubaneswar. The low flux of $1\times10^{12}$ ions/cm$^2$/s was used to avoid beam heating for all the irradiation. The focused beam was scanned over the area of 10 mm$\times$10 mm on the sample surface using electrostatic scanner for uniform irradiation. The projected range of 1 MeV Ni ions in MgO is 860 nm calculated from SRIM \cite{38}. The calculated electronic (S$_e$) and nuclear (S$_n$) energy loss is 1.06 keV/nm and 0.68 keV/nm for 1 MeV Ni on MgO, respectively. The energetic ions created a lot of vacancy and substitutional defects along their path, when 1 MeV Ni$^+$ ions were irradiated on MgO single crystal, lead to change in optical properties in the matrix. The absorption spectra were collected using UV-Vis spectroscopy in the wavelength range of 200-800 nm (Cary 5000 - Agilent). The variation of  bandgap  with ion fluences are procured from Tauc plotting. Steady state photoluminescence (PL)spectra were collected using a 325 nm of He-Cd (CW) laser at room temperature in the wavelength range of 350-630 nm. The sample were excited through an achromatic UV objective (LMU-UVB) with 10× magnification. The backscattered emission was collected through the same objective, using a CCD detector coupled to the spectrometer.

\section{\label{sec:level3}Computational Details}
The construction of defect bands have been calculated by taking MgO cubic super cell containing 32 formula units and inducing few vacancy and substitutional defects in it. The optimized crystal structure in the theoretical calculation for pristine, two Mg substituted by two Ni impurity, one Mg, one oxygen, four oxygen and two Mg and only four oxygen vacancy defect related MgO are shown in Fig. 1(a-f), respectively. The two Mg atoms replaced with two Ni atoms corresponds to 6.25$\%$ of Ni doping in MgO. The electronic structure calculation was done in the framework of DFT, implemented in the Vienna Ab initio Simulation Package (VASP) \cite{39,40}. Relaxation of the structures are done with generalized gradient approximation (GGA) scheme to consider the exchange correlation interaction with the Perdew, Burke, and Ernzerhof (PBE) functional \cite{41,42}. MgO has the cubic structure belongs to the space group of Fm$\overline3$m. The structural relaxations are completed when the force reaches the value smaller than 0.001 $eV/\AA$ for each atom. The most stable structures have been taken to calculate the band structure and density of states (DOS). The plane wave cut-off energy was fixed at 600 eV for pristine and vacancy impurity and 1000 eV for substitutional defect related structures. A suitable k points mesh of $12\times12\times12$ was used for reaching the required convergence results within $10^{-6}$ eV per atom. The GGA calculations are in good qualitative agreement with the experimental results. The bandgap of crystalline MgO in our calculation is 4.50 eV matching closely with literature value \cite{43,44,45}.

\begin{figure}[h!]
\includegraphics [width=8cm]{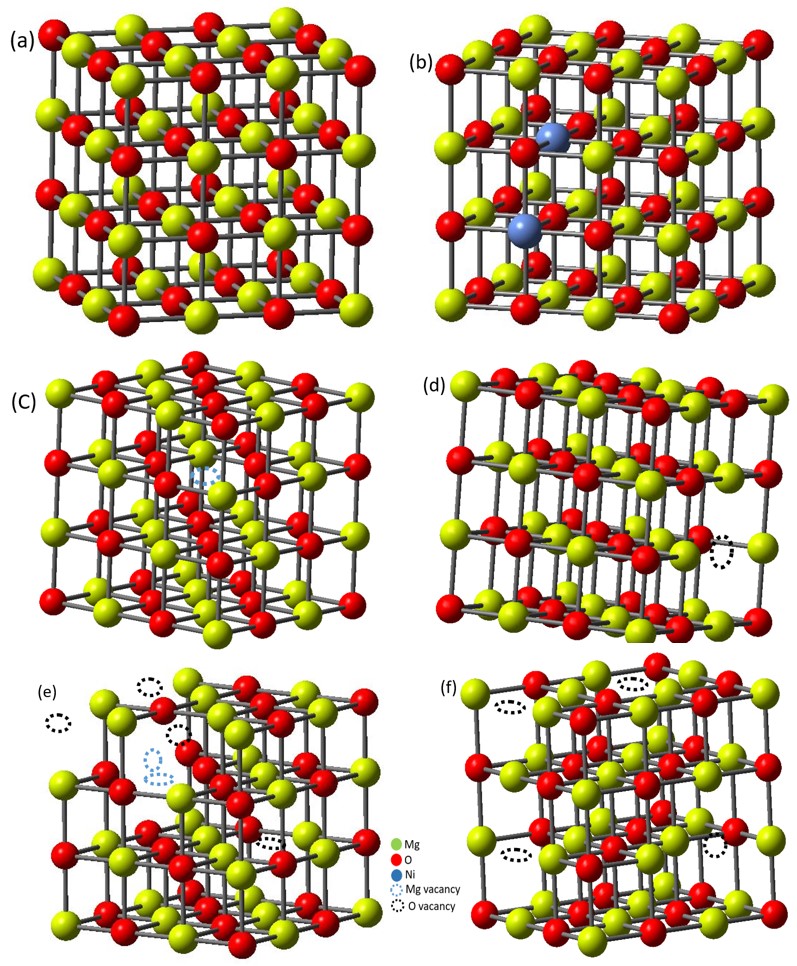}
\caption{\label{fig:picture_1} The atomic configuration of (a) pristine, (b) Ni impurity, (c) one Mg vacancy, (d) one oxygen vacancy, (e) four Oxygen and two Magnesium vacancy and (f) four oxygen vacancy defect MgO used for the theoretical calculation. Red, green and blue colour indicates the O, Mg and Ni atoms respectively.}
\end{figure}

\section{\label{sec:level4}Results and Discussions}

\subsection{UV-VIS ABSORPTION} Optical absorption spectra is a sensitive method to understand the defect centre and bandgap of a materials \cite{46}. The UV-Vis absorption spectra of pristine and MeV Ni$^{+}$ ion implanted MgO single crystal is shown in Fig. 2(a). It can observe from the figure that maximum absorption occurs near 247.8 nm (5.0 eV) for the fluence of $1\times10^{16}$ ions/cm$^2$ which was identified as F type defect centre \cite{47} that occurs due to oxygen vacancies containing two electrons. The integrated intensity of F type defects are increasing with ion fluences. The two week absorption peaks around 306 and 360 nm are due to the transition of oxygen defect centre which are assigned to $1A_{1g}\rightarrow1E $(electron donation from $1s \rightarrow 2p_x$ or $2p_y)$ and $1A_{1g}\rightarrow 1A_{1g} $(electron donation from $1s \rightarrow 2p_z$) respectively \cite{35}. The small peak of $F_2$ type defect \cite{37} centre formed at 360 nm in implanted sample is due to oxygen di-vacancy center. We also observed another defect centre called cationic or V centre \cite{4} (Mg vacancy centre) at 575 nm.

\begin{figure}[h!]
\includegraphics[width=\linewidth]{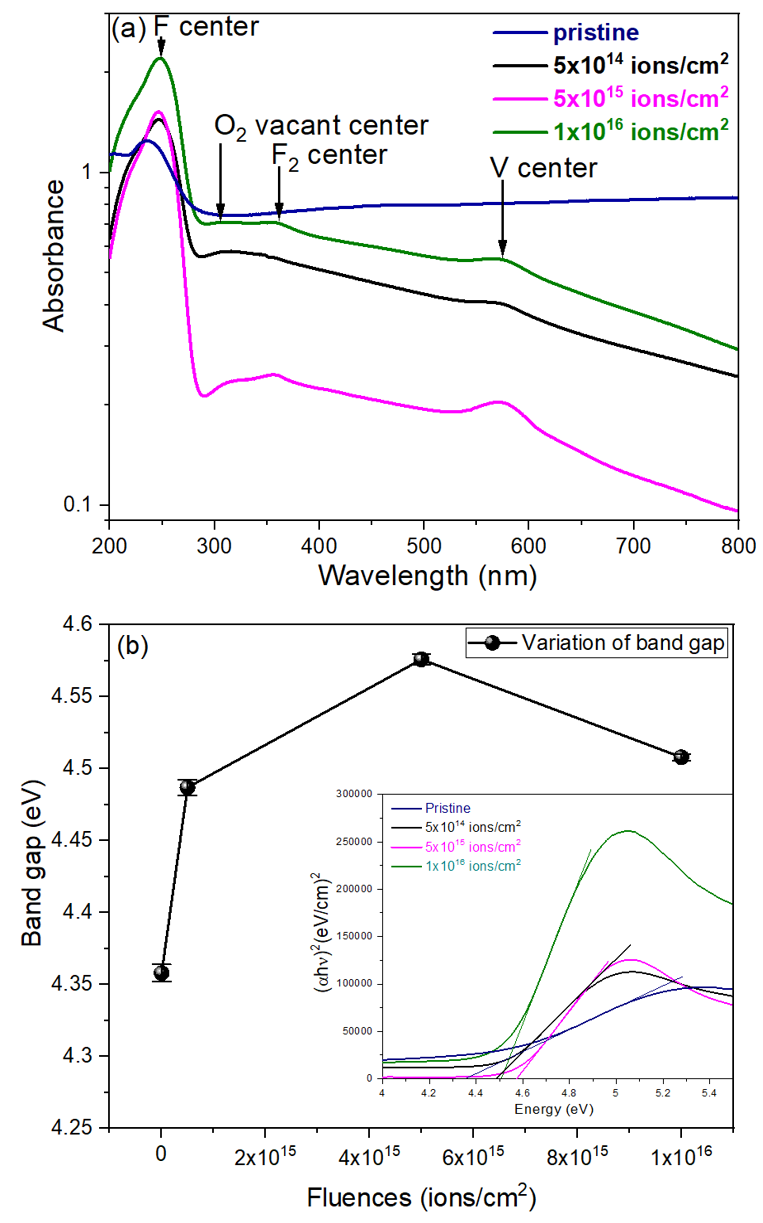}
\caption{\label{fig:picture_2}Absorption vs wavelength spectra of MeV N$i^+$ ions implanted MgO with different fluences shown in Fig. (a). Variation of optical bandgap of MgO with implanted Ni ion fluences from 0 to $1\times 10^{16}$ ions/cm$^2$ is shown in Fig. (b). The Tauc plotting i.e $(\alpha h\nu)^2$ vs energy $(h\nu)$ plotting of pristine and MeV Ni implanted MgO single crystal is shown in inset of Fig. (b).}
\end{figure}

There is no F$_2$ and V type defect centre present in pristine but MeV Ni ion induces these defect centres in implanted samples which become prominent with fluences as seen in the Fig. 2(a). When the high energetic Ni ions incident on crystalline MgO substrate, there is a possibility of substitution of Mg ions by Ni ions. Devaraja et al. mentioned that, the substitution is possible when the difference in percentage of radius (D$_r$) between the implanted and substituted ions does not exceed 30 \%. The radius percentage difference is calculated from the following formula \cite{50}:

\begin{equation}
    D_r = \frac{Rm(CN)-Rd(CN)}{Rm(CN)}
\end{equation}

Where, CN = Co-ordination number, Rm(CN) = Radius of host cation ($Mg^{2+}$), Rd(CN) = Radius of implanted ions ($Ni^{2+}$). We have found $D_r$ of 4.16 $\%$ which is much lower than 30 $\%$, which confirms Ni ions can substitute Mg ions. But it is observed that if the radius of the cation in the host lattice is much larger than the implanted ions, the implanted ions will always sit at interstitial site \cite{51}. This transition depends on relative ionic radius and relative oxygen affinity between the host and implanted cations. So it can be confirmed that the defect at 575 nm is due to Mg vacancy center. We can also say that Ni ion can't substitute the $O^{2-}$ ions because in this case, $D_r$ is high as the ionic radius of $O^{2-}$ is much larger than the $Ni^{2+}$ ions.
The optical bandgap of the pristine as well as Ni implanted MgO samples having fluence range from $5\times 10^{14}$ ions/cm$^2$ to $1\times 10^{16}$ ions/cm$^2$ is shown in Fig. 2(b). The bandgap is calculated from the intersect of the energy axis of the Tauc plot, shown in inset of Fig. 2(b) \cite{52}. The relation between absorption coefficient and incident photon energy used in Tauc plot follows as:

\begin{equation}
    (\alpha h \nu)^n = C(h\nu - E_g)
\end{equation}

Where, $\alpha$ = absorption coefficient, h = Planck's constant, $\nu$ = incident photon frequency, C is a constant, $E_g$ = energy gap and the value of n represents the nature of transition and MgO has found to be a direct bandgap nature. The optical bandgap of pristine is found at 4.36 eV. It is observed that the optical bandgap increases with the increase of Ni ion fluences and reached its maximum (4.58 eV) at a fluence of $5\times 10^{15}$ ions/cm$^2$. After that the bandgap decreases gradually and become 4.51 eV at the fluence of $1\times 10^{16}$ ions/cm$^2$. This kind of optical bandgap variation with Ni implanted MgO with different ion fluences observed from UV-Vis spectra is not reported extensively elsewhere. Mainly Burstein–Moss shift \cite{53} and bandgap narrowing phenomenon \cite{54} are the dominating phenomenon to effect optical bandgap \cite{55}. Bandgap can also vary due to change in crystal structure, lattice strain, surface and interface effect \cite{56}. But in this case, the initial increment of the optical bandgap with the increment of the incident Ni ions fluences may arise due to a) Ion implantation in MgO lattice defects change the lattice parameter in such a way that the movement of band edges towards the near band edges occurs due to the higher order of band bending results the enhancement of the bandgap.
b) When high energetic Ni ions are irradiated on the MgO single crystal, a huge amount of oxygen vacancy defects created as observed in absorption spectra. This leads to the higher density of oxygen defects in MgO and push the absorption band to higher energy resulting the increment of optical bandgap \cite{hari}.

The increment of optical bandgap continues up-to the fluence of $5\times 10^{15}$ ions/cm$^2$ and then decreases because a) The rises of oxygen vacancy tends to saturate with the increment of ion fluences. The creation of vacancies with further increment of Ni ion fluences over the recombination of vacancy fill-up becomes lower, resulting the decrease of oxygen defect in the matrix which causes the decreasing trend of optical bandgap.\\b) The exchange interaction between s, p and d orbitals of localised electrons of Ni ions, colour centre and band electrons may be the another reason for decreasing of bandgap after arising the maximum gap at a fluence of $5\times 10^{15}$ ions/cm$^2$ \cite{56}.

\subsection{PHOTOLUMINESCENCE SPECTRA}

The photoluminescenece (PL) spectra were studied to understand the type of defects and the quality of materials. Fig. 3(a) shows the room temperature PL spectra of pristine and Ni implanted MgO in the fluence range of $5\times 10^{14}$ ions/cm$^2$ to $1 \times 10^{16}$ ions/cm$^2$. Fig. 3(b) shows a typical PL spectrum corresponds to $5\times10^{14}$ ions/cm$^2$ sample, which is deconvoluted into four peaks at 387, 417, 518, and 565 nm.

\begin{figure}[h!]
\includegraphics[width=0.7 \linewidth]{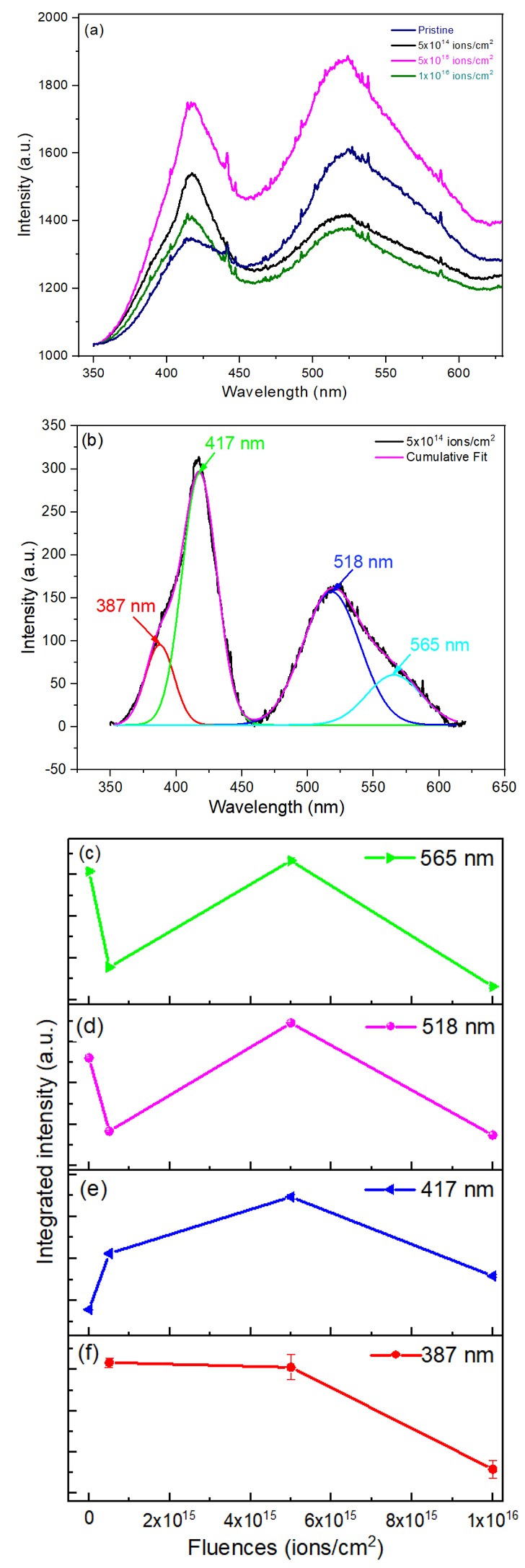}
\caption{\label{fig:picture_3}Photoluminescence spectra of pristine MgO and Ni implanted MgO at an excitation wavelength of 325 nm. The typical deconvoluted spectra for corresponding fluence (b) $5\times10^{14}$ $ions/cm^2$  is shown. The integrated intensity for (c) 565, (d) 518, (e) 417 and (f) 387 with different ion fluences found from deconvoluted spectra for pristine and implanted samples are presented.}
\end{figure}

The PL band peak detected around 565 nm after deconvolution of the spectra is associated with the oxygen vacancy center. This peak is identified as $F^-$ center shown in Fig. 3(c) \cite{extra}. It is already described about the formation of the vacancy defect centers elaborately in our previous section. Edel et al. \cite{65} reported the model and mechanism for F center luminescence of MgO. There are two ways to explain the luminescence behavior of the F center. The UV photon source stimulates the F center to $F^*$ and $F^+$ excited state. The $F^*$ state is de-excited by emitting a photon of 2.37 eV. The electron makes a non-radiative transition from later metastable $F^+$ state to radiative $F^*$ state and then de-excite by emitting photon energy of the same amount. The peaks approximately at 518 and 417 nm realized from Fig. 3(d) and 3(e) are identified as F and V type defect band center respectively \cite{65,66}. The decrease of intensity for both the bands at fluence $1\times 10^{16}$ ions/cm$^2$ may be due to the increase of recombination of F and V center. The intensity of 416 nm peak increases with ion fluences and then decreases with ion fluences after $5\times 10^{15}$ ions/cm$^2$. The PL peak at 416 nm arises due to the electronic transition from $F^7_1$ to $D^5_3$ \cite{67a}. It can be observed from Fig. 3(a) that there is a blue shift of the PL peaks, ensuring the increase of bandgap, and the redshift at the highest ion fluence tends to decrease the bandgap. This trend corroborates well with the variation of bandgap from UV-Vis absorption spectra.

\begin{figure}[h!]
\includegraphics[width=\linewidth]{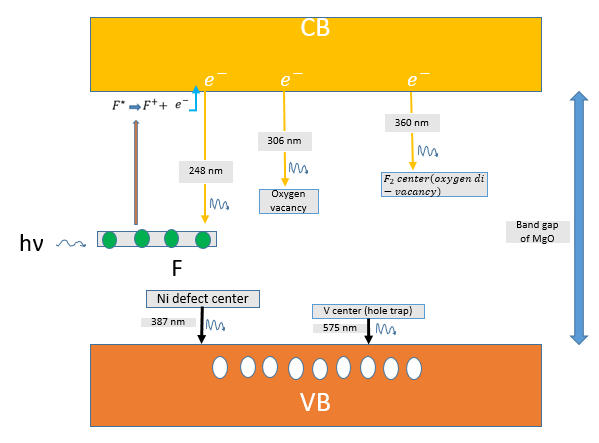}
\caption{\label{fig:picture_4} Defect model based on different possible free electronic transition. The circles in F centre represent the electronic state and the filled circle here represent that all the electronic states are filled by electrons. The white circle represents the hole in the valance band.}
\end{figure}

Fig. 4 describes the possible mechanism of the formation of the above defects found from absorption and PL spectra. When the incident light excites F type defect center, it de-excites by releasing one electron. The released electron can go to the conduction band (CB) and becomes free there. Now this CB electron can recombine with various defects states fromed between CB and valance band (VB). The CB electron de-excited to the F-center by releasing energy with a wavelength of 248 nm. The transition occurs towards another oxygen vacancy center or oxygen di-vacancy center, releasing energy having wavelength of 306 and 360 nm, respectively. The recombination of CB electron is not possible with V center \cite{34}. The hole trapped defect center lies near the VB maximum \cite{19}.However, there is a possibility of transition of the less excited electrons from VB to V-type defect center. Such de-excitation from the V center to VB leads to the emission of energy corresponding to the wavelength of 575 nm.  The metallic defect center near VB mainly formed by Ni-d orbital, confirmed from the band structure calculation. The less energetic electrons can jump to the nearest Ni defect center rather than CB, leaving a hole in the VB. When the electrons de-excite, the electron-hole recombination happens with VB holes by the emission of photon energy of wavelength 387 nm.

\section{\label{sec:level5}THEORETICAL RESULTS AND EXPLANATIONS}

\begin{figure}[h!]
\includegraphics[width=8cm]{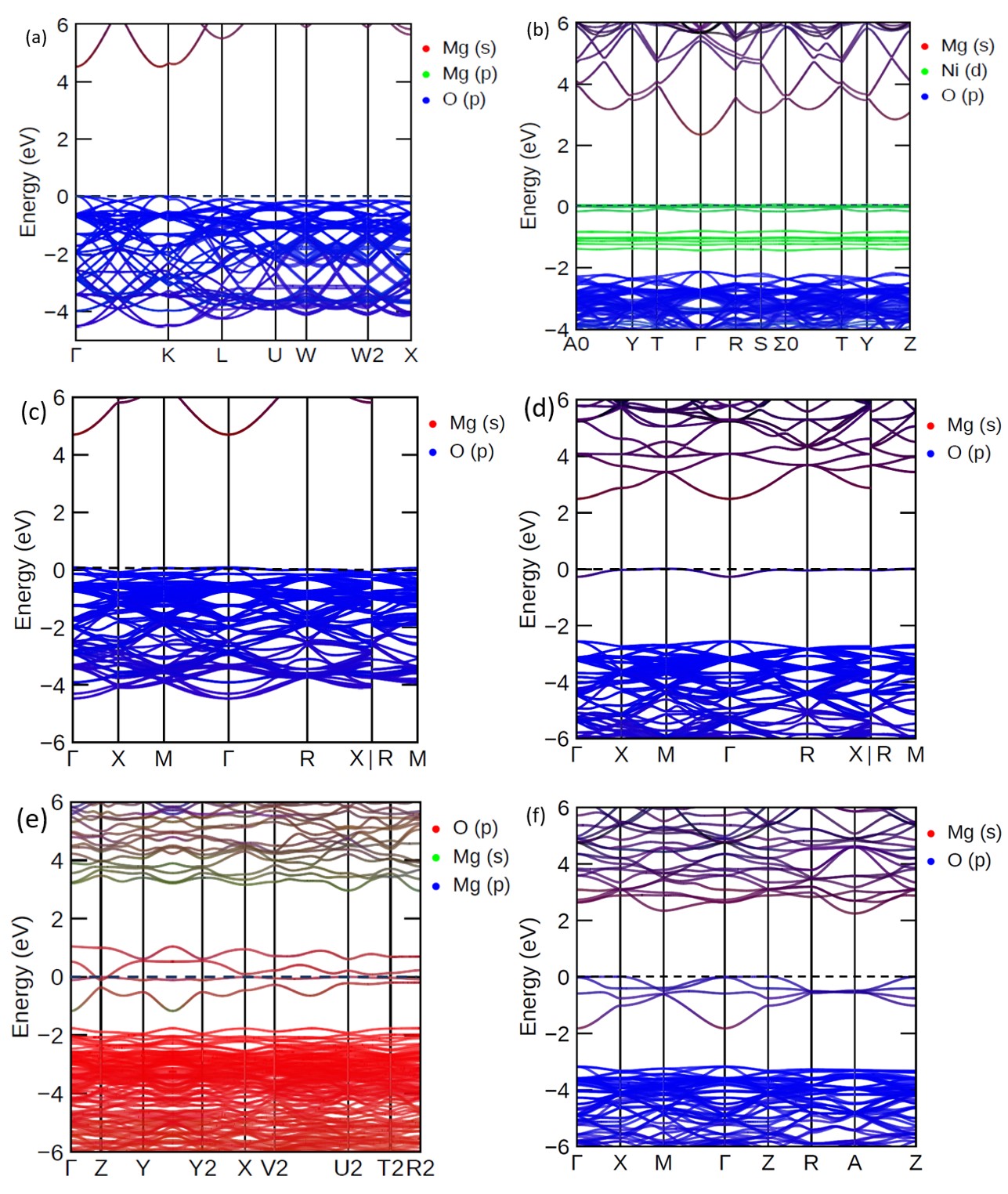}
\caption{\label{fig:picture_5}The calculated band structure of (a) pristine, (b) 2 Mg substituted by Ni (c) one Mg vacancy, (d) one oxygen vacancy, (e) two Mg and four oxygen vacancy and (f) four oxygen vacancy MgO structures. The dashed black line at 0 eV denotes the Fermi level. }
\end{figure}

\begin{figure}[h!]
\includegraphics [width=8cm]{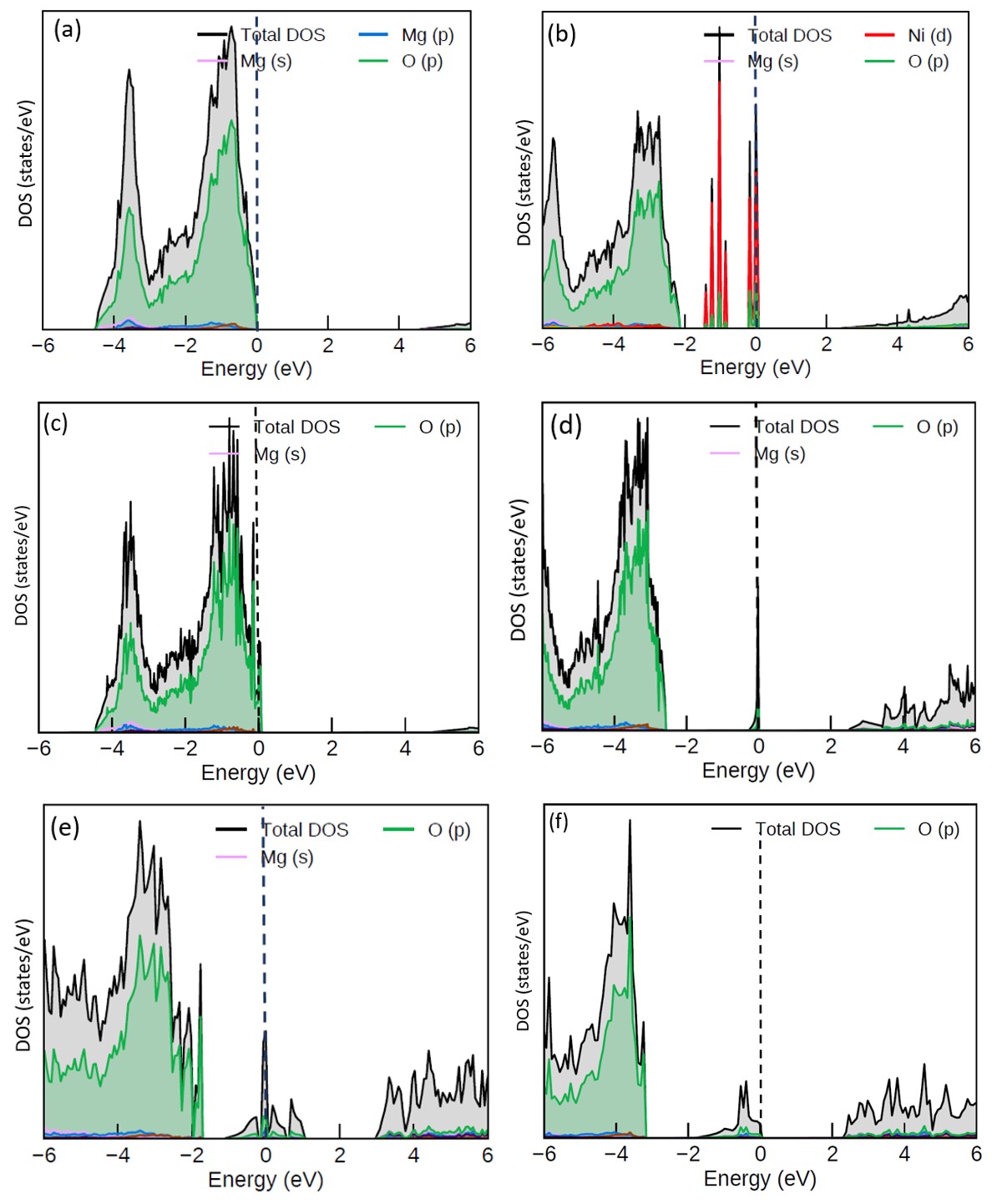}
\caption{\label{fig:picture_6}The calculated total and projected density of states (PDOS) of (a) pristine, (b) 2 Mg substituted by Ni, (c) one Mg vacancy, (d) one oxygen vacancy, (e) two Mg and four oxygen vacancy and (f) four oxygen vacancy MgO structures. The dashed black line at 0 eV indicates the Fermi level.  }
\end{figure}

The theoretical calculation of electronic band structures of pristine, substitutional defects, and vacancy defects in MgO are shown in Fig. 5. The band structure of pristine in Fig. 5(a) shows that the VB and CB mainly consist of O-p and Mg-s and Mg-p orbital, respectively. We found an energy gap of 4.50eV between the CB minimum and VB maximum for the pristine, which closely matches with our experimental band gap. Fig. 5(b) and 5(d-f) yield defect bands near the Fermi level between VB and CB appear after introducing substitutional, one oxygen, two oxygen and four Mg and four oxygen vacancy defect in pristine, respectively. So these types of defects influence the bandgap variation as seen in experimental results. Oxygen-p orbital has the most contribution to form the VB. The Mg-s, Mg-p, and O-p states have significantly contributed to create the CB for these substitutional and vacancy defect structures. Fig. 5(b) shows that when Ni substitutes Mg, the defect band around the Fermi level mainly consists of Ni-d orbital. The creation of Mg vacancy in MgO don't have critical role to create defect states near Fermi level whereas one oxygen vacancy creates a defect state there. This is confirmed from Fig. 5(c) and 5(d), respectively. Since the number of oxygen vacancy defects are proportional to the number of production of defect states (Fig. d and f), the large intensity for oxygen vacancy defects in absorption spectra confirms that most of the defect states are due to oxygen vacancy states. To justify the role of Mg vacancy with the combination of oxygen vacancy in formation of defect states and bandgap variation, we showed in Fig. 5(e) that the vacancy defect formed around Fermi level due to O-p orbitals only. Fig. 5(f) further justified the contribution of the four oxygen vacancy for the arising of four defect band and apprise that the Mg vacancy can only play a role in the movement of oxygen defect states around Fermi level to reduce the bandgap. The band structure calculation confirms the formation of different defect states between CB and VB, explaining the reason to take all the defect states between CB and VB in our proposed model. It is observed from Fig. 5 that the CB minimum and VB maximum occurs at the same gamma point, confirming the direct bandgap property of MgO. This DFT calculation also strongly supported the exactitude of choosing n=2 in Tauc plotting for pristine and implanted samples.
The DOS calculations further verified the formation of these substitutional and vacancy defect states near the Fermi level. Fig. 6 shows the total DOS and projected density of states (PDOS) of pristine and substitutional and vacancy defect induced MgO. It is seen from Fig. 6(b), 6(d-f) that the creation of substitutional and vacancy defect states is mainly due to Ni-d and O-p states, respectively. Fig. 6(c) further tells that Mg vacancy has no role to form the defect states near the Fermi level. Substituting two Mg by two Ni can produce more effective defect states than two Mg and four oxygen vacancies and only four oxygen in MgO, confirms by comparing Fig 6(b) and 6(e,f), respectively. This effect is greatly matched with our experimental results shown in Fig. 2. Since Mg has minimal contribution to form the defect bands, the intensity of the absorption peak near 575 nm is small. The major contribution of the vacancy defect comes from the O-p states and the transfer of these defects are influenced by Mg vacancy states, proved from Fig 5(e,f) and 6(e,f). This state also affects in the higher intensity absorption band peak near 247 nm. Similar phenomena can also be noticed from the band structure calculations. So these defect states can be filled fully or partially by electrons and become free after moving to the CB. These freed electrons can return to different defect states or jumps to VB directly or via defect states. The Fermi level was near the VB in pristine, then shifts towards the CB after defect formation, as seen in Figures 5 and 6, which is responsible for variation of bandgap. Since Ni-d, O-p state mainly form the defect states, electron exchange interaction between these states may influence the decreasing bandgap trend at a higher fluence of Ni ions. So, the bandgap variation with ion fluences is well justified from the band structure and DOS calculations. The deconvoluted PL band around 387 nm arises due to defect creation by high energetic Ni ions in MgO. Ni-d orbital mainly contributed to creating this defect band, as seen in Figures 5(b) and 6(b). So, the electronic band structure and DOS calculations confirmed the presence of Ni-induced defect bands in implanted samples.\\
It is interesting to note that the low fluence ion implantation increases the bandgap of MgO, i.e tending towards the higher insulating property. It can be utilized for the industrial cables, a mineral-insulated cable \cite{van}, for using in resistance of critical electrical circuit during fire. Such type of MgO system would be a suitable candidate for the fire protection device like fire alarm and smoke control devices. MgO is a potential binary metal oxide candidate which can be applied as an insulating layer in transition metal oxide (TMO) based resistive random access memories (RRAM) due to it's large bandgap property \cite{loy}. In this report, the tuning of bandgap of MgO with ion fluences can impart many crucial idea for making various type of better RRAM. The domination of conduction mechanism in MgO memory device depends on the hopping and Ohmic conduction in high resistance state (HRS) and low resistance state (LRS), respectively. But the hopping conduction is related to energy level from trap state to the CB. It is determined that the trap states in MgO films come from the F centers, trapping center for electron \cite{chiu}. We showed here that most of the vacancy defect are due to oxygen vacancy and the intensity of this vacancy defects are much higher than the other defects, formed in the matrix. So, this anionic vacancies can be applied to form the filament in valance charge memory (VCM) based RRAM.

\section{\label{sec:level6}Conclusion}
In conclusion, we have investigated the creation of different kinds of defects in crystalline MgO after 1 MeV Ni ion implantation and observed the variation of optical bandgap as a function of ion fluences. The UV-Vis absorption spectra demonstrate that the absorption at the F center is more prominent than the V center. A small amount of absorption at the oxygen di-vacancy center is also observed. The optical bandgap increases monotonically with ion fluences till a fluence of $5\times 10^{15}$ ions/cm$^2$ and then decreases. We understand the nature of the bandgap of MgO from the DFT calculation and showed that the enhancement of Fermi level towards the CB and the electron exchange interaction between Ni-d, O-p, and Mg-s states are responsible for the variation of the bandgap. The excited electrons become free after moving from the F center to the CB via $F^+$ center.The electrons are de-excited to different defect states by releasing photon energy. This mechanism aids us in identifying various defect centers observed in absorption spectra. The observation of F and V center in PL spectra further confirms in identifying defect centers. In addition the application of the vacancy defects to form the the filament in VCM and the utilities of this kind of bandgap variation in TMO based RRAM device is justified. 

\begin{acknowledgments}

The authors acknowledge the National Institute of Science Education and Research Bhubaneswar, DAE, India for supporting this work. Sourav Bhakta and PKS acknowledge the stafff of ion beam facility at Institute of Physics Bhubaneswar for providing stable beam for ion implantation.  SB acknowledge the  scientific officer Mr. A Ananda Raman for his help to use the Kalinga cluster of NISER.  

\end{acknowledgments}

\bibliography{apssamp}% 

\end{document}